\documentclass[12pt]{article}

\newcommand{\ad}{a^{\dagger}}

\newcommand{\be}{\begin{equation}}
\newcommand{\bea}{\begin{eqnarray}}
\newcommand{\ba}{\begin{array}}
\newcommand{\ee}{\end{equation}}
\newcommand{\eea}{\end{eqnarray}}
\newcommand{\ea}{\end{array}}
\newcommand{\bean}{\begin{eqnarray*}}
\newcommand{\eean}{\end{eqnarray*}}
 \newcommand{\ft}[2]{{\textstyle {\frac{#1}{#2}} }}
\newcommand{\RR}{{\mathbb R}}

\usepackage{amsfonts, latexsym}

\newcommand{\n}[1]{{:}\, #1 {:} }
\newcommand{\e}[1]{\langle \, #1 \, \rangle}

\begin{document}

\thispagestyle{empty}

\begin{flushright}
{\tt physics/0212061}
\end{flushright}

\vspace{2cm}

\begin{center}
{\bf\Large Wick Calculus}
\medskip

\bigskip\bigskip

{\bf 
Alexander Wurm\footnote{\tt alex@einstein.ph.utexas.edu}
 and Marcus Berg\footnote{\tt mberg@physics.utexas.edu}
}

\vspace{.3cm}  
 Center for Relativity  and 
Department of Physics \\
The
University of Texas at Austin\\ 
Austin, TX 78712

\renewcommand{\thefootnote}{\arabic{footnote}}
\setcounter{footnote}{0}
\bigskip
\medskip

\begin{abstract}
In quantum field theory, 
physicists routinely use
``normal ordering'' of operators, 
which just amounts to shuffling all creation operators to the left.
Potentially confusing, then,
is the occurrence in the literature
of normal-ordered {\it functions},
sometimes called ``Wick transforms''.  
We aim to introduce the reader to some 
basic results and ideas around this theme, 
if not the mathematical subtleties; 
our intended audience are 
instructors who want to add something to their 
quantum field theory course, or researchers who 
are interested  but not specialists in mathematical physics.
For rigorous proofs and careful mathematical discussions we
only give references.
\end{abstract}
\end{center}

\newpage

\section{Introduction}
Normal ordering was introduced in quantum field theory
by G.C. Wick in 1950, 
to avoid some infinities in the vacuum 
expectation values of field operators expressed in terms of creation
and annihilation operators \cite{Wick}.
The simplest example of such infinities 
can be discussed based on non-relativistic quantum mechanics and  
the simple harmonic oscillator;
an infinite number of harmonic oscillators make up a free quantum field. 
(The reader who needs a quick reminder of some 
basic quantum field-theoretical concepts may
find comfort in appendix A). We will 
use the harmonic oscillator to exhibit the connection between
Wick-ordered 
polynomials and the familiar Hermite
polynomials. Then we turn to Wick transforms in the functional integral
formalism of field theory, where we show that there is again a connection
with Hermite polynomials. Several different approaches
to Wick transforms that can be found in the literature 
are compared and we show 
why they are equivalent.  In passing, we observe how the standard
quantum field theory result known as ``Wick's theorem'' 
follows rather directly in this 
framework, from well-known properties of Hermite polynomials. 
Finally, we provide one brief
example of how the Wick transform can be utilized in a
physical application.

\section{Wick  operator ordering}

\subsection{Simple Harmonic Oscillator}

\noindent
The Hamiltonian operator for the simple harmonic oscillator in
non-relativistic quantum mechanics has the form
\[
H = \ft12 (P^2 +Q^2)  \; ,
\] 
where we have, as usual, hidden
Planck's constant $\hbar$, the mass $m$, and the angular frequency 
$\omega$ in 
definitions of dimensionless momentum and position operators
\[
P:=\frac{1}{\sqrt{\hbar m \omega}} \, \hat{p}
\; ,\qquad\qquad Q:=\sqrt{\frac{m\omega}{\hbar}} \, \hat{q} \; ,
\]
so that
\[
[P,Q] = -i \; .
\]
If one defines the creation and annihilation operators
$\ad$ and $a$ by
\begin{eqnarray}
\ad & = & \frac{1}{\sqrt{2}} (Q - iP) \; ,\\
a & = & \frac{1}{\sqrt{2}} (Q+iP) \; , 
\end{eqnarray}
so that
\be
[a, \ad] =1 \; ,
\ee
one finds that
\be
H = \ft12 (P^2 +Q^2) = \ft12 \left( \ad a+a\ad\right) =
\ad a +\ft12 \quad . \label{ham}
\ee
This means,
as proved in introductory quantum mechanics books (see e.g.
Sakurai \cite{Sakurai}), that the
eigenvalues of the Hamiltonian operator 
come in the sequence
\[
E_n = n +\frac{1}{2} \; , \qquad n = 0, 1, 2,\ldots \; .
\]
In particular,
the ground state energy (or zero-point energy), which is the lowest 
eigenvalue of the Hamiltonian, is non-zero:
\[
H |0\rangle = \ft12 \,   |0\rangle.
\]
This is in agreement with Heisenberg's uncertainty principle: It is the 
smallest energy value that saturates 
the uncertainty relation
(again, see e.g. Sakurai \cite{Sakurai} for the explicit calculation).

This zero-point energy has observable physical 
consequences;
as an illustration, 
it is possible to measure zero-point motion (which leads
to the zero-point energy) of atoms in a crystal by studying
dispersion of light in the crystal. Classical theory 
predicts that any oscillations
of the atoms in the crystal, and therefore also dispersion effects,
 cease to exist when the temperature is lowered towards absolute zero.
 However, experiments demonstrate that dispersion of light reaches
a finite, non-zero value at very low temperature.

In quantum field theory, a free scalar field can be viewed as an infinite
collection of harmonic oscillators, as described in appendix A. If we 
proceed as before, each oscillator will give a contribution to the zero-point
energy, resulting in an infinite energy, which seems like
it could be a problem.

One way to remedy the situation is to {\it define} the ground state as a state
of zero energy. 
We  can achieve this by redefining the Hamiltonian: We subtract the contribution
of the ground state and define a so-called {\it Wick-ordered} 
(or normal-ordered) 
Hamiltonian, denoted by putting a colon on each side, by
\be
\n{H} = \n{\ft{1}{2}\left(\ad a + a\ad\right)} \stackrel{\rm def}{=}
\ft{1}{2} \left( \ad a +a \ad\right) - \ft{1}{2} \langle 0| \ad a
+ a \ad |0\rangle \equiv \ad a.
\label{WickH}
\ee
Hence, in this example the definition of Wick ordering
can be thought of as a redefinition of the ground state
of the harmonic oscillator.

On the other hand, we also see that in the last equality in 
(\ref{WickH}) all creation operators end up on the left.
A general prescription for Wick ordering 
in quantum field theory in a creation/annihilation operator formalism
is then:
``Permute all the $\ad$ and $a$, treating them as if they commute,
 so that in the end all $\ad$ are
to the left of all $a$.''  The resulting expression is, of course, the same:
\[
\n{H} = \ad a \; .
\]

\subsection{Wick ordering and Hermite polynomials}

The first connection between Wick ordering and Hermite polynomials arises
if we study powers of the (dimensionless) position operator $Q$.
Physically, in the harmonic oscillator, the eigenvalue
of $Q^2$ 
gives the variance (squared standard deviation) of the oscillator from rest.
We have $Q=(\ad +a)/\sqrt{2}$, but to 
avoid cluttering the equations with factors of $\sqrt{2}$, we will 
study powers of just $(\ad +a)$.
\bean
(\ad + a)^2 &=& a^{\dagger 2} + \ad a + a \ad + a^2 \\
            &=& a^{\dagger 2} + 2\ad a + a^2 + [a,\ad]\\
            &=& \n{(\ad +a)^2} + [a, \ad] \\
            &=& \n{(\ad +a)^2} + 1 \qquad\qquad\qquad\qquad \mbox{by (5)}. 
\eean
Arranging terms in a similar way for higher powers of $(\ad + a)$ we find
\bean
(\ad + a)^3 &=& :(\ad + a)^3: + \, 3(\ad + a) \; .  \\
(\ad + a)^4 
            & = & :(\ad + a)^4 : + \, 6 :(\ad + a)^2: + \, 3 \; .
\eean
We can summarize the results as follows, with 
the notation $\ad +a =q$,
\bean
q^2 & = & \n{q^2} + 1 \; , \\
q^3 & = & \n{q^3} + \, 3\n{q} \; , \\
q^4 & = & \n{q^4} + \, 6 \n{q^2} +\, 3 \; .
\eean
Since we can  recursively replace normal-ordered terms on the
right by expressions on the left which are not normal-ordered 
(e.g. $\n{q^2}$ 
can be replaced by $q^2-1$), we can also invert these relations: 
\[
\ba{rcll}
:q^2: & = & q^2 - 1 & = {\rm He}_2(q) \; , \\
:q^3: & = & q^3 - 3q & = {\rm He}_3(q) \; , \\
:q^4: & = &  q^4 - 6q^2 + 3 & = {\rm He}_4(q)\; ,\\
\ea
\]
where 
the polynomials ${\rm He}_n(q)$ are a scaled version of the more 
familiar form of the
Hermite polynomials ${\rm H}_n$:
\be
{\rm He}_n(x)=2^{-n/2} {\rm H}_n(x/\sqrt{2}).
\label{eq:scaled}
\ee
Confusingly, in some mathematical physics
literature, the ${\rm He}_n$ are often just called ${\rm H}_n$.
Some of the many useful properties
are collected in appendix B for easy reference
(a more complete collection is given in e.g.\
Abramowitz \& Stegun \cite{AbrSte}). 

Because of this relation between operator Wick-ordering
and Hermite polynomials,
the mathematical physics literature
sometimes {\it defines} ``Wick ordering'' in terms of Hermite polynomials:
\be
\n{q^n} \stackrel{\rm def}{=} {\rm He}_n(q)\,.
\label{seven}
\ee
Although $q$ is an operator composed of noncommuting operators
$a$ and $\ad$, this alternative definition 
naturally generalizes to Wick-ordering of functions.
As promised, we will explore this idea in the next section.

One reason that the connection to Hermite polynomials
is not mentioned in standard quantum field theory 
literature is the fact (which was also Wick's motivation) that
the normal-ordered part is precisely 
the part that will vanish when one takes
the vacuum expectation value. 
Indeed, the traditional way to define normal ordering,
the one given
at the end of section 2.1 (``put $\ad$ to the left of $a$''), yields
for powers of $q$ simply
\[
\n{q^n} = \sum_{i=1}^n \left(\ba{c} n \\ i\ea \right)\; 
(a^{\dagger})^{n-i} 
a^{i},
\]
which vanishes for any nonzero power $n$
when applied to the vacuum state $ | 0\rangle$.

In other words, since one knows that normal ordered
terms vanish upon taking the vacuum expectation
value, one may not be interested in their precise  
form. 

However, when the expectation value 
is not taken in the vacuum (for example, in a particle-scattering
experiment), this part does of course not vanish in general,
and there are in fact many instances where the actual normal-ordered
expression itself is the one of interest.

\section{Functional integrals and the Wick transform}

Most modern courses on quantum field theory
discuss functional integrals 
(sometimes called ``path'' integrals, however 
only in nonrelativistic quantum mechanics does one really integrate
over paths). 
In a functional-integral setting,
the counterpart of the Wick ordering in the operator formalism
is the {\it Wick transform}.
This transform applies to functions and functionals. It can, like 
its quantum-mechanics
counterpart eq.\ (\ref{seven}), be defined by means of Hermite polynomials.
But first, let us briefly skip ahead and explain why 
such a transform will prove to 
be useful.

\subsection{Integration over products of fields}

In the functional formalism, physical
quantities like scattering cross sections and decay constants
are computed by integrating over some polynomial in the fields
and their derivatives. 
The algorithmic craft of such computations is described in textbooks such 
as Ryder \cite{Ryder} and Peskin \& Schroeder \cite{PesSch}.
Even though there are
examples of physical effects
that can be studied with functional-integral
methods but not with ordinary canonical quantization\footnote{In
particular, contributions to the $S$-matrix that
have an essential singularity at zero coupling constant
cannot be found by standard perturbative expansion around zero coupling,
yet these ``nonperturbative'' contributions
can be studied using functional integrals.
Rates for decays that would 
be strictly forbidden without these effects can be computed,
see for example the book by Ryder \cite[Ch.\ 10.5]{Ryder}.}, 
within the scope of this paper we can only
give examples of some things that can be derived more quickly or
transparently using functional integrals.

In fact, we will also be concerned with some rather basic questions
that are usually glossed over in introductory treatments:
what does the functional integral itself really mean? While
a complete answer is not even known, and certainly
beyond the scope of this short article, we intend to give some
flavor of the first steps towards addressing this question and
how the Wick transform has been put to work in this regard. 

First, a restriction:
the polynomials considered in this section are polynomials of Euclidean
fields (fields defined on 
four-dimensional Euclidean space $\RR^4$).
Similar formalisms exist for 
Minkowski fields (fields defined on spacetime) with minor changes
in the equations (see e.g. the aforementioned textbook \cite{PesSch}). 
Unfortunately, functional integrals over 
Minkowski fields are less mathematically developed 
than integrals over Euclidean
fields, therefore  we shall restrict attention to Wick transforms of 
functions and functionals of Euclidean fields --- primarily polynomials
and exponentials. The Wick transform, like
the Hermite polynomials, has orthogonality properties that turn out
to be useful in quantum field theory, as we shall see.
First, we have to introduce a few mathematical
concepts.

\subsubsection{Gaussian measures}
Here, our aim is to fix the notation, and to briefly
remind the reader how
to integrate over Euclidean fields, without going into too much detail. 
The standard mathematical framework to perform such integration
is the theory of
 Gaussian measures in Euclidean field theory, for which 
details can be found in
the mathematical physics literature, such as
Glimm \& Jaffe
 \cite{GliJaf} and Janson  
\cite{Janson}.

As a first try, one would define a field
in the functional integral as a function $\phi$ on $\RR^4$.
Fields in the functional integral, however, may seem like 
functions at first glance, but
can produce divergences
that cannot (for instance) be {\it multiplied} 
in the way that functions can. 
A more useful way to regard
a quantum field in the functional integral formalism is
as a distribution
 $\Phi$ acting on a space of test functions $f$:
\be
\Phi(f)\stackrel{\rm def}{=} \langle\Phi , f\rangle \; ,
\ee
where the bracket $\langle \; , \; \rangle$ denotes
duality, i.e.\ $\Phi$ is such that it yields a number when applied
to a smooth test function $f$.
In many situations the distribution $\Phi$ is equivalent to a
function $\phi$, which means this number is the ordinary integral
\be
\langle \Phi, f\rangle=\int_{\RR^4} d^4x \; \phi(x) f(x) \; .
\label{distrib}
\ee
A familiar example of a distribution
is the one-dimensional Dirac distribution $\Phi=\delta$, 
for which we have
\[
\Phi(f)=\langle \delta, f \rangle = f(0) \; .
\]

Now, just like a function, $\Phi$ in general 
belongs to an infinite-dimensional space.
To be able to integrate over this space
(not to be confused with the integral in equation (\ref{distrib}),
which is an ordinary integral over spacetime)
 we need a measure, some
generalization of the familiar
$dx$ in the ordinary integral above. 
Here, a useful generalization will actually depend on 
the Green's function, called 
{\it covariance} in this context and denoted $C$.
 In general, 
the covariance is a positive, continuous, 
non-degenerate bilinear form on the space of test functions.
In the following,
we will often encounter the covariance at coincident test functions,
here denoted $C(f,f)$.
% This is usually a divergent
%constant, but by using its inverse as normalization 
%it can be disposed of, and later on  we will set it to unity. 

To get to the point,
a Gaussian measure $d\mu_C$ is defined by its covariance $C$ as
\be
\int_{\bf Y} d\mu_C (\Phi)\; \exp \left( -i \langle\Phi, f\rangle\right) = 
\exp \left( - \ft12 C(f,f) \right)
\label{eq:C}
\ee
over a space {\bf Y} of distributions $\Phi$.
%, or
%\be
%\int_Y d\mu_C(\phi) \; \exp 
%\left(- i \langle\phi', \phi \rangle \right) = \exp 
%\left( -\ft12 \langle\phi', C\phi\rangle\right).
%\label{eq:C2}
%\ee
%In eq.\ (\ref{eq:C2}),
%the duality 
%$\langle\phi', \phi\rangle$ is determined by the space
%$Y$ of functions  $\phi$,
%generalizing the standard $\RR^d$ Gaussian measure
%to spaces whose dual is not necessarily
%identifiable with itself. 
For comparison, the usual Gaussian measure on 
$\RR^d$ is defined by
\footnote{
 To avoid dimension-dependent numerical
terms (powers of $\pi$, powers of $2$) in the definition (\ref{finite})
of the Gaussian measure one can, alternatively, define it by
\[
\int_{\RR^d} {d^dx \over  (\det D)^{1/2}} \; 
e^{- \pi Q(x)} e^{-2\pi i \langle x',x\rangle} = e^{ - \pi
W(x')} .
\]
In fact, this can even be convenient in the simplest Fourier transforms
for those who forget where the $2\pi$ goes:
$\hat{f}(p)=\int dx \; e^{-2\pi i p x} f(x)$ yields an inverse
$f(x)=\int dp \; e^{2\pi i p x} \hat{f}(p)$, without forefactor.
}
\bea
&&{1 \over (2\pi)^{d/2}} 
\int_{\RR^d}{d^dx \over (\det D)^{1/2}} \; e^{-
\ft12 Q(x)}\; e^{-i \langle x',x\rangle}
= e^{ - \ft12 W(x')}
\label{finite}
\eea
where 
\[
 \langle x',x\rangle_{\RR^d}\; =  x_{\mu}' x^{\mu}\;\;\; 
\mbox{is the duality in $\RR^d$},
\;\;\; x\in \RR^d,\;
 x'\in \RR_d,
\]
\[
Q(x)\;\;\; \mbox{is a quadratic form on}\; \RR^d,\;\;\;
Q(x) =D_{\mu\nu} x^{\mu} x^{\nu} = \langle 
Dx, x\rangle_{\RR^d},
\]
\[
W(x')\;\;\;\mbox{is a quadratic form on}\; \RR_d,\;\;\;
W(x') = x_{\mu}' C^{\mu\nu} x_{\nu}' = \langle
x', C x'\rangle_{\RR^d},
\]
such that
\be
D C = C D = 1 \; .
\label{CD}
\ee
In the more familiar $\RR^d$ case (\ref{finite}), 
it is the {\it combination} of standard measure
and kinetic term that corresponds to
the measure $d\mu_C$ we introduced above:
\be
{d^d x \over (\det D)^{1/2}}\; e^{-\ft12 Q(x)} \qquad \mbox{is
analogous to}
\qquad d\mu_C(\Phi) \; ,
\ee
but going back from $d\mu_C$ to an explicit separation 
as on the left
will not turn out to be necessary 
for our discussion. In fact, by
defining the measure $d\mu_C$ through eq.\ (\ref{eq:C}),
we have not even specified what such a separation would mean.

With the above expressions in mind, the covariance at incident points 
is expressed as the following integral, 
obtained by expanding eq.\ (\ref{eq:C}):
\be
C(f,f) = \int d\mu_C(\Phi)\; \langle\Phi, f\rangle^2  \; .
\label{sigma}
\ee
In fact, the integral
on the left-hand side of eq.\ (\ref{eq:C}) is the generating
function of the Gaussian measure; let us denote this integral by $Z(f)$. 
This means that by successive expansion of eq.\ (\ref{eq:C}),
the $n$-th moment of the Gaussian
measure can be compactly written as
\bea
\int d\mu_C(\Phi)\; \langle\Phi ,f\rangle^n & = &
 \left( -i \frac{d}{d\lambda}\right)^n \!\!
Z(\lambda f)\; {\Bigg |}_{\lambda =0}\nonumber\\[.25in]
& = & \left\{\begin{array}{cl}
(n-1)!!\,\, C(f,f)^{n/2} & n\;\; \mbox{even} \\[2mm]
 0 & n\;\;  \mbox{odd, }
\end{array}\right.
\label{moments}
\eea
where $n!!\, =\, n (n-2) (n-4)\cdots$ is the semifactorial.
For convenience, we introduce the following
notation for the average with respect to the
Gaussian measure $\mu_C$:
\be
\langle\,  F[\Phi( f)] \, \rangle_{\mu_C} 
:= \int d\mu_C(\Phi)\; F[\Phi(f)] \; .
\ee
Note the difference between the brackets $\langle \quad \rangle_{\mu_C}$
used for average and the brackets 
$\langle\; , \; \rangle$ used for duality.
Armed with this set of definitions, we can 
define a Wick transform of functionals of fields.

\subsubsection{Wick transforms, definitions}

The goal here is to provide some idea of how to address
the difficult mathematical
problem of making sense out of products of distributions, and
integrals of such products, which (as argued above)
are ubiquitous in quantum field theory, although
their exact meaning is not usually discussed in standard
introductory textbooks.
In order to simplify quantum field theory calculations, 
one defines the Wick
 transform of a power $\Phi(f)^n \, := \, 
 \langle\Phi^n,  f\rangle$ 
so as to
 satisfy an orthogonality property with respect to  Gaussian integration.
Recalling the orthogonality properties of Hermite polynomials (appendix B)
and the definition (\ref{seven}) a simple idea is 
to {\it define} the Wick transform in terms of Hermite 
polynomials:
\be  \fbox{$\displaystyle
\n{\Phi(f)^n}_C = C(f,f)^{n/2}{\rm He}_n
\!\left({\Phi(f) \over \sqrt{C(f,f)}}\right). $}
\label{wherm}
\ee
Notice that this depends on the covariance $C$, and that
there is no analogous dependence in
the analogous harmonic-oscillator definition
(\ref{seven}).

The orthogonality of two 
Wick-transformed polynomials is then expressed by
\be
\int d\mu_C(\Phi)\; \n{\Phi(f)^n}_C\; \n{\Phi(g)^m}_C\; =
\; \delta_{m,n}\; 
n!\; (\langle\Phi(f)\; \Phi(g)\rangle_{\mu_C})^n.
\ee
An entertaining exercise is to show this, which we will do 
in section 3.1.3 (paragraph 2).

The Wick transform can also be defined recursively by the
following equations \cite{Simon}
\begin{center}
{\it Wick transform defined recursively}
\end{center}
\be \fbox{$\displaystyle
\ba{rcll}
\n{\Phi(f)^0}_C &=&1 & \\[2mm]
{\displaystyle {\delta \over \delta \Phi}}
\n{\Phi(f)^n}_C &=&n \; \n{\Phi(f)^{n-1}}_C \qquad & 
n = 1,2,... \\[3mm]
\int d\mu_C(\Phi)\;\n{\Phi(f)^n}_C &=& 0 & n= 1,2,... \\[2mm]
\ea $}\label{recurdef}
\ee
where the functional derivative with respect to a distribution
is simply
\[
{\delta\over\delta\Phi} \Phi(f) = f.
\]
Let us check that the Wick transform $\n{\Phi(f)^n}_C$
defined by
eq.\ (\ref{recurdef}) is the same as the Wick transform
 given in terms
of Hermite polynomials in eq.\ (\ref{wherm}). 
This is of course to be expected, since Hermite
polynomials themselves satisfy 
similar recursion relations, but it
is a useful exercise to check that it works.
To begin with, we
establish a property of Wick exponentials.

\begin{center}
{\it Wick exponentials}
\end{center}

Let $\n{\exp(\alpha\Phi(f))}_C$ be the formal series
\[
\n{\exp(\alpha\Phi(f))}_C \equiv 1+\alpha\n{\Phi(f)}_C+\ft12 \alpha^2
\n{\Phi(f)^2}_C + ...
\]
where normal-ordering is defined by eq.\ (\ref{recurdef}). \\
{\it Exercise:} Show that
\be
\n{\exp(\alpha\Phi(f))}_C ={\exp(\alpha \Phi(f)) \over 
\; \e{\exp(\alpha 
\Phi(f))}_{\mu_C}} \; .
\label{eq:exp}
\ee

\noindent
{\it Solution:}\\
One can evaluate  the right-hand side of the equation
by expanding numerator and denominator into a power series and dividing
one power series by the other.
\[
\frac{\sum_{k=0}^{\infty} b_k\, x^k}{\sum_{k=0}^{\infty} a_k\, x^k} =
\frac{1}{a_0} \sum_{k=0}^{\infty} c_k\, x^k
\]
where $c_n +\frac{1}{a_0}\sum_{k=1}^n c_{n-k}\, a_k -b_n =0$.
Comparing the resulting series, term by term, to the power series expansion
of the left side proves equation (\ref{eq:exp}).

\begin{center}
{\it Equivalence of Hermite polynomial and recursive definitions}
\end{center}

\noindent
{\it Exercise:} Show the equivalence of 
(\ref{recurdef}) and (\ref{wherm}). 

\noindent
{\it Solution:}
We can explicitly calculate the denominator in eq.\ (\ref{eq:exp}):
\bean
\e{\exp(\alpha \Phi(f))}_{\mu_C}&=&\int d\mu_C(\Phi) \, \exp(\alpha \Phi(f)) \\
&=& \int d\mu_C(\Phi)\, \sum_n \frac{\alpha^n}{n!}\Phi(f)^n \\
&=& \sum_n \frac{\alpha^{2n}}{n!} \int d\mu_C(\Phi) \;\Phi(f)^{2n}
\qquad\qquad\mbox{by eq.\ (\ref{moments})}\\ 
&=& \exp\left(\ft12 \, \alpha^2 C(f,f)\right)\qquad\qquad
\qquad\quad\mbox{by eq.\ (\ref{sigma})}. 
\eean
Thus, from eq.\ (\ref{eq:exp}) we find
\be
\n{\exp(\alpha \Phi(f))}_C=\exp\left(\alpha \Phi(f) - \ft12 \, \alpha^2
 C(f,f)\right).
\label{exposeries}
\ee
Multiplying the power series expansions\footnote{Similarly 
to the division of power series
mentioned above, the multiplication of power series
is simply
\[ \left( \sum_{k=0}^{\infty} a_k\, x^k\right) \left( \sum_{k=0}^{\infty}
 b_k\, x^k\right)
= \sum_{n=0}^{\infty} d_n\, x^n
\]
\indent
where $d_n =\sum_{m=0}^n a_m\, b_{n-m}$.}
of $\exp(\alpha \Phi(f))$
 and $\exp(\frac{1}{2}\alpha^2 C(f,f))$ and comparing the 
result term by
term to the series expansion of the left side of eq.\ (\ref{exposeries})
yields
\be
\n{\Phi(f)^n}_C = \sum_{m=0}^{\left[ \frac{n}{2}\right]} \frac{n!}{m!\, 
(n-2m)!} \Phi(f)^{n-2m} \left( -\ft12 C(f,f)\right)^m.
\label{explic}
\ee
Rewriting this expression as
\be
\n{\Phi(f)^n}_C = C(f,f)^{n/2}\;
\sum_{m=0}^{\left[ \frac{n}{2}\right]} (-1)^m
 \frac{n!}{2^m m!\, (n-2m)!} \left(\frac{\Phi(f)}{\sqrt{C(f,f)}})
\right)^{n-2m},
\label{explic2}
\ee
and using the formula for the defining series of the Hermite polynomials
given in appendix B, one recovers eq.\ (\ref{wherm}).

\subsubsection{Wick transforms, properties}
\label{sec:prop}

Many properties of Wick ordered polynomials can be conveniently derived
using the formal exponential series. For simplicity, we assume
that
all physical quantities that one 
may wish to compute (scattering cross sections, etc.) 
are written with normalization factors of
 $1/C(f,g)$, which in effect lets us set
the coincendent-point 
covariance to unity: $C(f,f)=1$. It can be restored by comparison
 with eq.\ (\ref{explic2}). 
The  properties we are interested in are useful exercises
to show:

\begin{itemize}
\item[1.] 
Show that \\
$\n{\exp\left(\Phi(f)+\Phi(g)\right)}_C = \exp\left(-\e{\Phi(f)
\,\Phi(g)}_{\mu_C}\right)\,
 \n{\exp(\Phi(f))}_C\,\n{\exp(\Phi(g)}_C$

\noindent
{\it Solution:}
\begin{eqnarray*}
&&\n{\exp(\Phi(f))}_C\, \n{\exp(\Phi(g)}_C \\
& & \qquad\qquad =  \exp\left(\Phi(f)+\Phi(g)
\right)\, 
 \exp\left(-\frac{1}{2}\left[\e{\Phi(f)^2}_{\mu_C}\, +\,
 \e{\Phi(g)^2}_{\mu_C}\right]\right)\\
& & \qquad\qquad \n{\exp\left(\Phi(f)+\Phi(g)\right)}_C\, \exp\left(\e{\Phi(f)
\,\Phi(g)}_{\mu_C}\right),
\end{eqnarray*}
where we have used eq.\ (\ref{exposeries}) in the first line, and, after
completing the square in the second factor, again in the second line.
Dividing both sides of the equation by the second factor completes the 
demonstration.

\item[2.] Show that \\
$\e{\n{\Phi(f)^n}_C\, \n{\Phi(g)^m}_C}_{\mu_C} = \delta_{n\,m}\, n!\,
 \e{\Phi(f)\, \Phi(g)}^n_{\mu_C} $

\noindent
{\it Solution:}\\
 If we take the expectation value of both sides of the last line in the 
proof above, we find
\[
\langle \n{\exp(\Phi(f))}_C\, \n{\exp(\Phi(g))}_C \rangle_{\mu_C} =
 \exp\left(\e{\Phi(f)\,
 \Phi(g)}_{\mu_C}\right)
\]
using eq.\ (\ref{recurdef}). Expanding the exponentials on both sides and
comparing term by term completes the proof.

\item[3.] Show that \\
$\n{\Phi(f)^{n+1}}_C = n
 \n{\Phi(f)^{n-1}}_C -\Phi(f)\, \n{\Phi(f)^n}_C$

\noindent
{\it Solution:}\\
This is a  consequence of the equivalence of Wick ordered
functions and Hermite polynomials. The expression follows from the recursion
relation for Hermite polynomials given in appendix B.

\item[4.] The definition of Wick transforms  given above can be
generalized to several
fields in a very straightforward manner. We quote here some results without 
proof (details can be found in e.g.\ the book by Simon \cite{Simon}).
The reader may find it interesting to check that it works: 
\begin{eqnarray*}
&&\n{\Phi(f_1) \ldots \Phi(f_{n+1})}  =  \n{\Phi(f_1)\ldots \Phi(f_n)}\, 
\Phi(f_{n+1})\\ & &\qquad\qquad  - \sum_{k=1}^n C(f_k , f_{n+1})\, \n{\Phi(f_1)\ldots
\Phi(f_{k-1}) \Phi(f_{k+1})\ldots \Phi(f_n)}
\end{eqnarray*}
\[
\int d\mu_C(\Phi)\, \n{\Phi(f_1)\ldots \Phi(f_n)} = 0
\]
\[
\int d\mu_C(\Phi)\, \n{\Phi(f_1)\ldots \Phi(f_n)}\, \n{\Phi(g_1)\ldots
\Phi(g_m)}=0  \qquad\qquad \mbox{for}\; n\neq m
\]
\end{itemize}

These latter multi-field expressions reproduce,
within this functional framework, what is usually referred to as
``Wick's theorem'' in the creation/annihilation-operator formalism. 
In that formalism, it takes some effort to show this theorem; here 
we find it somewhat easier, relying on
familiar properties of the Hermite polynomials. 

\subsection{Wick transforms and functional Laplacians}

\subsubsection{Definition}

We can also define Wick transforms of functions by the following
exponential operator expression, which is convenient in many cases 
(e.g. in two-dimensional quantum field theory settings, such as 
in references \cite{Polchinski}  and \cite{ItzDro}):
\be 
\fbox{$\displaystyle \n{\phi^n(x)}_C \stackrel{\rm def}{=} e^{-\frac{1}{2}
 \Delta_C}\; \phi^n(x) $ }
\label{lapl}
\ee
where the functional Laplacian is defined by 
\be
\Delta_C = \int d^4 x \, d^4 x' \, C(x,x')
{\delta \over \delta \phi(x)}
{\delta \over \delta \phi(x')} \; .
\ee
which, again, depends on the covariance $C$.

Instead of proving eq.\ ({\ref{lapl}), 
which is straightforward, we will 
just illustrate the equivalence
of definition (\ref{lapl}) and definition (\ref{wherm})
in the common example of a $\phi^4$ power, when the definition reads:
\[
\n{\phi^4(y)}_C=\exp\left(-{1 \over 2}\int d^4 x \, d^4 x' \, C(x,x')
{\delta \over \delta \phi(x)}
{\delta \over \delta \phi(x')}\right) \phi^4(y) \; .
\]
Expanding the exponential we find
\bean
\n{\phi^4(y)}_C&=&\phi^4+\left(-{1 \over 2}\right)
\left(\int d^4 x \, d^4 x' \, C(x,x')
{\delta \over \delta \phi(x)}
{\delta \over \delta \phi(x')}\right)\,\phi^4(y) \\
& & +
{1 \over 2!}
\left(-{1 \over 2}\right)^2\left(\int d^4 x \, d^4 x' \, C(x,x')
{\delta \over \delta \phi(x)}
{\delta \over \delta \phi(x')}\right)^2\phi^4(y) \; .
\eean
All higher terms in the expansion are zero. 
We can now evaluate each term
separately;
the second term is the integral
\bean
&&\int d^4 x \, d^4 x' \, C(x,x')
{\delta \over \delta \phi(x)}
{\delta \over \delta \phi(x')}\,\phi^4(y) \qquad \qquad \\
\quad &= & 
\int d^4 x \, d^4 x' \, C(x,x')
{\delta \over \delta \phi(x)}\,4\phi^3(y)\, \delta(x'-y) \\
& =& \int d^4 x \, C(x,y) \, 12 \phi^2(y)
\, \delta(x-y)  \\
& =& 12 \phi^2(y)
\eean
if the covariance is normalized to unity.  
We use this result in the third term:
\bean
&&\int d^4 x \, d^4 x' \, C(x,x')
{\delta \over \delta \phi(x)}
{\delta \over \delta \phi(x')} \, 12 \phi^2(y)\\
& &\qquad\qquad = \int d^4 x \, d^4 x' \, C(x,x')
{\delta \over \delta \phi(x)} \, 24 \phi(y) \delta(x'-y) \\
& &\qquad\qquad = 24 \int d^4 x  \, C(x,y)
 \, \delta(x-y) \\
& &\qquad\qquad = 24
\eean
with the same normalization of the covariance. We collect these results with
the appropriate coefficients from the expansion:
\bean
\n{\phi^4(y)}_C&=&
\phi^4(y)-{1 \over 2}\cdot 12\,\phi^2(y)
+{1 \over 2!}{1 \over 2^2} \cdot 24\\[1mm]
&=&\phi^4(y)-6\, \phi^2(y)+3 \\[2mm]
&=&{\rm He}_4(\phi(y)) \; ,
\eean
which completes the illustrative example.

\subsection{Further reading}

Although we hope to have given some flavor of
some of the techniques and ideas of quantum field theory 
mathematical-physics style, 
we have of course really only given a few illustrative
examples and demonstrated some simple identities.
For more on the mathematical connection between
Wick transforms on function spaces
and Wick ordering of annihilation and creation operators,
we recommend textbooks such as
\cite{GliJaf} and \cite{Janson}.\\

\subsection{An application: specific heat}

In this last section,
we discuss one example of a physical application of 
some of the results above.
By using the 
connection (\ref{wherm})
between the Wick transform and Hermite polynomials, we 
show how one can 
exploit standard properties of those polynomials
to simplify certain calculations.

Consider the familiar generating function of Hermite polynomials (but
for the scaled polynomials (\ref{eq:scaled})):
\be
e^{x\alpha-\alpha^2/2} =\sum {\rm He}_n(x)\; \alpha^n/n! \quad .
\label{genH}
\ee
This generating function
gives a shortcut to computing some quantum effects 
in two-dimensional quantum field theory, 
where normal orderering is often the only form of renormalization
necessary.
An important issue is the scaling dimension of
the normal-ordered exponential $\n{e^{ip\phi}}\, $,
where $p$ is a momentum. (The real part of this operator can represent the energy
of a system where $\phi$ is the quantum field.). 
In other words, the question is: 
if we rescale our momentum $p \rightarrow \Lambda p$, 
equivalent to a rescaling $x \rightarrow \Lambda^{-1}x $ in coordinate
space, how
does this operator scale? Since an exponential 
is normally
expected to be dimensionless, one might guess that the answer is 
that it does not scale at all, i.e. that the scaling dimension is
zero.
In fact, this is not so due to quantum effects
induced by the normal ordering.
Passing to the functional integral, we can easily compute
the effect of the normal ordering (here, the Wick transform):
\bea
\n{e^{ip\phi}}_C & = & \sum_{n=0}^{\infty} \frac{1}{n!} (ip)^n
\n{\phi^n}_C  \nonumber \\
& = & \sum_{n=0}^{\infty} \frac{1}{n!} (ip)^n C^{n/2} {\rm He}_n
\left(\phi/\sqrt{C}\right)\nonumber \\
& = & e^{\frac{1}{2}p^2 C} e^{ip\phi} \; ,
\label{anom}
\eea
where we used the definition (\ref{wherm}) and the previously given
generating function (\ref{genH}). 
Now, the covariance (Green's function) $C$
is a logarithm in two dimensions, i.e.\ the solution 
of the two-dimensional Laplace equation is a logarithm.  
To regulate divergences when $p\rightarrow \infty$, one introduces 
a cutoff $\Lambda$ on the momentum, which 
makes $C= \ln \Lambda $.
Substituting this into (\ref{anom}) yields the answer
\be
\n{e^{ip\phi}} = \Lambda^{p^2/2} e^{ip\phi}\; .
\label{andim}
\ee
Thus, the anomalous scaling dimension,
usually denoted by $\gamma$, is $\gamma=p^2/2$
for the exponential operator. This is an
important basic result in conformal field theory (see e.g.\ \cite{Witten},
p.\ 451). Here, the $p^2/2$ just comes from 
the $\alpha^2/2$ in the generating function (\ref{genH})!

How could such a quantum effect be measured?
Consider the "two-dimensional Ising model with random bonds" (p.\ 719, 
\cite{ItzDro}). 
This is just the familiar Ising model, but one allows the coupling
between spins to fluctuate, i.e.\ the coupling becomes a space-dependent
Euclidean field in the spirit of previous sections.
The energy of the system is described by (the real part of)
an exponential operator $\n{e^{ip\phi}}$ as stated above.
Using renormalization group methods, the anomalous dimension 
(\ref{andim}) leads to a
formula for the specific heat in this system
(eq.\ (356) in
\cite{ItzDro}, where the derivation is also given).
The specific heat is in principle directly measurable as a function of the
temperature, or more conveniently, as a function of  $\theta =
(T-T_c)/T_c$, the
dimensionless deviation from the critical temperature. 
The renormalization group description predicts a certain double
logarithm dependence on $\theta$ that could not have been found by
simple perturbation theory, and it uses
as input the result (\ref{andim}).

Admittedly, the telegraphic description in the previous
paragraph does
not do justice to the full calculation of the
specific heat in the two-dimensional Ising model
with random bonds. Our purpose here was simply to show how
the Wick transform reproduces the quantum effect (\ref{andim}),
and then to give some flavor of how this effect is measurable.

\section{Conclusion}

In this article we have shown that the scope of normal
ordering has expanded to settings beyond the original one
of ordering operators. Several different
definitions of Wick ordering of functions have been discussed and their
equivalence established. 

For deeper understanding and further applications
of these ideas, the interested reader is invited to 
consult the quoted literature, which 
is a selection of texts we found particularly useful.
Specifically, for the physics of functional integrals we enjoy \cite{GliJaf}. 
For a more mathematically oriented treatment
we find \cite{Janson} quite useful.

\bigskip
\bigskip
\newpage

\noindent
{\large\bf Acknowledgments }\\

\noindent
The authors would like to thank C. DeWitt-Morette, P. Cartier and
D. Brydges for many helpful discussions,
and M. Haack for comments on the manuscript.

\appendix

\section{Wick ordering of operators in QFT}

In this appendix, we briefly remind the 
reader how the need for Wick ordering
arises in the operator formulation of quantum field theory.
All of this is  standard material and can be found in
any introductory book on quantum field theory (e.g.\cite{PesSch}),
albeit in lengthier and more thorough form.
We set $\hbar =1$ throughout.

Consider a 
real scalar field $\phi (t, {\bf x})$ of mass $m$ defined at 
all points of
 four-dimensional Minkowski spacetime and satisfying the  Klein-Gordon
equation
\[
\left( \frac{\partial^2}{\partial\, t^2}-\nabla^2  + m^2\right) 
\phi(t,{\bf x}) = 0.
\]
The differential operator in the
parenthesis is one instance of what we call $D$ in the text.
The classical Hamiltonian of this scalar field is
\be
H = \ft12 \sum_{\bf x} \, \big[ (\pi(t,{\bf x}))^2 + 
(\nabla \phi(t,{\bf x}))^2 + m^2 \phi^2(t,{\bf x}) \big]
\label{Hclass}
\ee
where $\pi$ is the variable canonically conjugate  to $\phi$,
in fact it is simply $\pi = \partial \phi/ \partial t$.
Here we can think of the first term as the kinetic energy,
and the second as the shear energy.
This classical system is quantized in the canonical quantization scheme by treating
the field $\phi$ as an operator, and imposing equal-time commutation relations
\begin{eqnarray*}
\left[\phi (t, \mathbf{x}), \phi (t, \mathbf{x'})\right] &=&  0 \;, \\
\left[\pi (t, \mathbf{x}), \pi (t, \mathbf{x'})\right] &=&  0 \; , \\
\left[\phi (t, \mathbf{x}), \pi (t, \mathbf{x'})\right] &=& i \delta^{3}
(\mathbf{x}-\mathbf{x'}) \; ,
\end{eqnarray*}
The plane-wave solutions of
the Klein-Gordon equation are known as the field modes, $u_{\bf k}
(t, {\bf x})$. Together with their respective complex conjugates $u^*_{\bf k}
 (t, {\bf x})$
they  form a complete orthonormal basis, so the field $\phi$ can
be expanded as
\[
\phi (t, {\bf x}) = \sum_{\bf k} [ a_{\bf k}\, u_{\bf k} (t, {\bf x}) +
  \ad_{\bf k} \, u^*_{\bf k} (t, {\bf x})].
\]
The equal time commutation relations for $\phi$ and $\pi$ are then equivalent
to
\begin{eqnarray*}
\big[ a_{\mathbf{k}}\, , a_{\mathbf{k'}}\big] &=& 0 \; , \\
\big[\ad_{\mathbf{k}}\, , \ad_{\mathbf{k'}}\big] &=& 0 \; , \\
\big[ a_{\mathbf{k}} \, ,  \ad_{\mathbf{k'}}\big] &=&
\delta_{\mathbf{k\, k'}} \; .
\end{eqnarray*}
These operators are defined on a Fock space, which is a Hilbert space
made of $n$-particle states ($n= 0,1,\ldots$). The normalized basis
 ket vectors, denoted by $|\; \rangle$, can be constructed
starting 
from the vector $| 0\rangle$, the vacuum. The vacuum state $|0\rangle$ has
the property that it is annihilated by all the $a_{\bf k}$ operators:
\[
a_{\bf k} |0\rangle = 0 \, ,\qquad \forall\, {\bf k}.
\]
In terms of the frequency $\omega_{\bf k} = c\sqrt{|{\bf k}|^2+m^2}$,
the Hamiltonian operator
obtained from (\ref{Hclass}) is
\[
\hat{H} = \ft12 \sum_{\bf k} ( \ad_{\bf k} a_{\bf k} + a_{\bf k} \ad_{\bf k})
\omega_{\bf k} = 
\sum_{\bf k} (\ad_{\bf k} a_{\bf k} + \ft12 )\,  \omega_{\bf k} \; ,
\]
where  in the last step we used the commutation relations from above.
Calculating the vacuum energy reveals a potential problem:
\[
\langle 0| \hat{H} | 0\rangle = \langle 0|0\rangle\, \sum_{\bf k} \ft12
\, \omega_{\bf k} = \sum_{\bf k} \ft12 \, \omega_{\bf k} \; \longrightarrow \; 
\infty \; ,
\]
where we have used the normalization condition $\langle 0|0\rangle =
1$. This infinite constant can be removed as described in the text.

Propagation amplitudes in quantum field theory
(and hence scattering cross sections and decay
constants) 
are given in terms of expectation values
of time-ordered products of field operators. 
These time-ordered products  arise in the interaction Hamiltonian of an
interacting quantum field theory. The goal is to compute 
propagation amplitudes for
these interactions using essentially time-dependent perturbation theory,
familiar from quantum mechanics. At leading order in the coupling constant,
these products can be simplified, and the zero-point constant energy removed
by using Wick's theorem.

The way Wick ordering is applied in practice to calculations in QFT is
through ``Wick's theorem'', which gives a decomposition
of time-ordered products of field operators into sums of normal-ordered
products of field operators (again, we refer to 
e.g.\ \cite{PesSch}). In this paper, in 
section \ref{sec:prop},  
Wick's theorem appears in the
functional-integral formulation of the theory.

\section{Properties of ${\rm He}_n(x)$}

Here we list a few 
useful properties of the scaled Hermite polynomials. More can be
found in \cite{AbrSte}.
\ \\

\noindent
Defining series:
\[
{\rm He}_n(x) =  \sum_{m=0}^{\left[ \frac{n}{2}\right]} (-1)^m 
\frac{n!}{m!\, 2^m\, (n-2m)!}\, x^{n-2m} \; ,
\]
where $[n/2]$ is the integer part of $n/2$.

\noindent
Orthogonality:
\[
\int_{-\infty}^{\infty} dx\; e^{-x^2/2}\;{\rm He}_n(x)\, {\rm He}_m
(x) 
= \delta_{m\, n}
\sqrt{2\pi}\, n! \; .
\]
Generating function:
\[
\exp\left(x\alpha -\ft12 \, \alpha^2\right) = \sum_{n=0}^{\infty} 
{\rm He}_n(x)\, \frac{\alpha^n}{n!} \; .
\]
Recursion relation:
\[
{\rm He}_{n+1}(x) = x\, {\rm He}_n(x) - n\, {\rm He}_{n-1}(x) \; .
\]
The first five:
\bean
{\rm He}_0(x) & = & 1  \; , \\
{\rm He}_1(x) & = & x  \; , \\
{\rm He}_2(x) & = & x^2 -1 \; , \\
{\rm He}_3(x) & = & x^3 -3x  \; , \\
{\rm He}_4(x) & = & x^4-6x^2+3 \; .
\eean


\begin{thebibliography}{99}
\bibitem{Wick}
G.C. Wick, ``The evaluation of the collision matrix'', 
Phys. Rev. {\bf 80}, 268-272 (1950).
\bibitem{Sakurai}
 J. J. Sakurai, {\it Modern Quantum Mechanics} 
(Addison-Wesley, Reading, MA, 1994), rev. ed.
\bibitem{AbrSte}
 M. Abramovitz and I. A. Stegun, {\it Handbook of Mathematical Functions} (Dover, New York, 1965).
\bibitem{Simon}
B. Simon, {\it The $P(\phi)_2$ Euclidean (Quantum) Field Theory} 
(Princeton University Press, Princeton, N.J., 1974). 
\bibitem{PesSch}
 M. Peskin and D. V. Schroeder, {\it An Introduction to Quantum Field
Theory} (Addison-Wesley, Reading, MA, 1995). See in particular the 
chapter on functional integrals in quantum field theory.
\bibitem{GliJaf}
 J. Glimme and A. Jaffe, {\it Quantum Physics} (Springer-Verlag, New York, 1981), 2nd ed.
\bibitem{Janson}
S. Janson, {\it Gaussian Hilbert Spaces} (Cambridge University Press,
Cambridge U.K., 1997).
\bibitem{Polchinski}
 J. Polchinski, {\it String Theory, Volume I: An Introduction to the
Bosonic String} (Cambridge University Press, Cambridge, 1998).
\bibitem{Ryder}
L. H.\ Ryder, {\it Quantum Field Theory}, 
(Cambridge University Press, Cambridge, 1985).
\bibitem{Salmhofer}
 M. Salmhofer, {\it Renormalization: An Introduction}
 (Springer-Verlag, Berlin, 1999).
\bibitem{Witten}
 E. Witten,  ``Perturbative Quantum Field Theory'', in {\it Quantum
Fields and Strings: A Course for Mathematicians, Vol. 1}, P. Deligne et al.,
eds. (American Mathematical Society, 1999).
\bibitem{ItzDro}
 C. Itzykson and J.-M. Drouffe, {\it Statistical Field
Theory}, 2 vols. (Cambridge University Press, Cambridge, 1989).
\end{thebibliography}
\end{document}